\documentclass[12pt,a4paper]{article}

\usepackage{amsmath}
\usepackage{amssymb}
\usepackage{graphicx}
\usepackage{dsfont}
\usepackage{overpic}
\usepackage{cite}

\let\oldappendix=\appendix
\let\oldsection=\section
\renewcommand{\appendix}{\oldappendix%
\def\theequation{\Alph{section}.\arabic{equation}}%
\renewcommand{\section}{\setcounter{equation}{0}\oldsection}}

\newcommand{\beq}{\begin{equation}}
\newcommand{\eeq}{\end{equation}}
\newcommand{\beqa}{\begin{eqnarray}}
\newcommand{\eeqa}{\end{eqnarray}}
\newcommand{\no}{\nonumber}

\newcommand{\tr}{\mbox{tr}}

\newcommand{\newop}[2]{\def#1{\mathop{\mathrm{#2}}\nolimits}}
\newop{\artanh}{artanh}
\newop{\det}{det}
\newop{\tr}{tr}
\newop{\diag}{diag}
\newop{\Re}{Re}
\newop{\Im}{Im}

\begin{document}

\hfill 

\hfill 

\bigskip\bigskip

\begin{center}

{{\Large\bf  Greens function of a free massive\\ scalar field on the lattice}}

\end{center}

\vspace{.4in}

\begin{center}
{\large B.~Borasoy\footnote{email: borasoy@itkp.uni-bonn.de} and
        H.~Krebs\footnote{email: hkrebs@itkp.uni-bonn.de}}

\bigskip

\bigskip

Helmholtz-Institut f\"ur Strahlen- und Kernphysik (Theorie), \\
Universit\"at Bonn, \\ 
Nu{\ss}allee 14-16, D-53115 Bonn, Germany \\

\vspace{.2in}

\end{center}

\vspace{.7in}

\thispagestyle{empty} 

\begin{abstract}
We propose a method to calculate the Greens function of a free massive 
scalar field on the lattice numerically to very high precision. 
For masses $m < 2$ (in lattice units) the massive Greens function can be 
expressed recursively in terms of
the massless Greens function and just two additional mass-independent constants.
\end{abstract}\bigskip

\begin{center}
\begin{tabular}{ll}
\textbf{PACS:}& 12.90.+b \\[6pt]
\textbf{Keywords:}& lattice field theories, lattice Greens function, coordinate-space methods.
\end{tabular}
\end{center}


\vspace{.7in}

\vfill

\newpage
Lattice regularization offers a convenient tool to study non-perturbative 
features of field theories. The Lagrangian is formulated on a discrete
space-time lattice and the corresponding finite path integral is evaluated 
numerically.
In most instances, one is ultimately interested in the continuum limit
by decreasing the lattice spacing to zero, but for very small lattice spacings
the lattice path integral is plagued with ultraviolet divergences rendering
numerical lattice simulations impossible.

Perturbation theory plays an important role in providing the missing link between
numerical simulations and the desired physical continuum limit. Perturbative
calculations on the lattice are performed, e.g., to obtain renormalization constants
of lattice operators or non-universal coefficients of $\beta$-functions.
If the physical situation contains separate scales,
they are also necessary in order to match the perturbative short distance physics
with the non-perturbative long distance physics.

However, perturbative calculations on the lattice tend to be difficult,
since the Feynman integrals are rather complicated functions of the involved momenta.
In this respect, coordinate-space methods have been proven useful in the evaluation
of Feynman diagrams and allow a very precise determination of the continuum limit
of two- and even higher-loop
integrals \cite{LW}. In \cite{LW} this technique was applied to massless propagators.
The massive case was addressed in \cite{M}, but for space-time dimension less than
four, and explicit results for the Greens functions were obtained only in one and 
two dimensions.
In the present investigation, we extend the work of L\"uscher and Weisz \cite{LW}
to massive scalar fields by presenting an efficient method to calculate the associated Greens function
to very high precision.

We will first generalize the recursion relation for the Greens function
presented in \cite{LW} to massive propagators.
The free propagator for a massive scalar field in position space reads in lattice units
\beq
G(n) = \int_{-\pi}^\pi  \frac{d^4k}{(2 \pi)^4} \ \frac{ e ^{ik \cdot n}}{\hat{k}^2 + m^2} \ ,
\eeq
where $n=(n_1, n_2, n_3, n_4)$ and
\beq
\hat{k}^2 = \sum_{\mu=1}^4 \hat{k}_\mu^2 = 4 \sum_{\mu=1}^4 \sin^2 (k_\mu/2) \ .
\eeq
We define the forward and backward lattice derivatives
\beq
\nabla_\mu G(n) = G(n+\mu) - G(n) \ , \qquad \nabla_\mu^* G(n) = G(n) - G(n-\mu) 
\eeq
which yield the lattice Laplacian
\beq
\Delta = \sum_{\mu=1}^4 \nabla_\mu^* \nabla_\mu
\eeq
with
\beq
\Delta G(n) = \sum_{\mu=1}^4 \Big( G(n+\mu) + G(n-\mu) - 2 G(n)  \Big) \ .
\eeq
The Greens function satisfies the equation
\beq
\Big( -\Delta + m^2  \Big) G(n) = \delta_{n0} 
\eeq
from which one obtains after making use of hypercubic symmetry the identity
\beq \label{greendiff}
\Big( 8 + m^2 \Big) G(0) - 8 G(\mu) = 1
\eeq
for any $\mu$.
The central identity utilized in \cite{LW} is given by
\beq
\Big( \nabla_\mu^* + \nabla_\mu  \Big) G(n) = n_\mu H(n) \ ,
\eeq
where
\beq
H(n) = \int_{-\pi}^\pi  \frac{d^4k}{(2 \pi)^4} \ e ^{ik \cdot n}   
          \ln \Big(\hat{k}^2 + m^2 \Big)
\eeq
is independent of $\mu$.
By eliminating $H(n)$ via
\beq
H(n) = \frac{2}{\rho} \sum_{\mu=1}^4  \Big( \Big[1+ \frac{m^2}{8} \Big] G(n) - G(n-\mu)  \Big)
\eeq
for $\rho = \sum_{\mu=1}^4 n_\mu \ne 0$
one obtains a recursion relation for the Greens function at different lattice sites
\beq \label{recursion}
G(n+\mu) = G(n-\mu) + \frac{2 n_\mu}{\rho} \sum_{\nu=1}^4 
             \Big( \Big[1+ \frac{m^2}{8} \Big] G(n) - G(n-\nu)  \Big) .
\eeq
This recursion relation is at the heart of this method, as it allows to express
the Greens function at any lattice site in terms of a few values close 
to the origin. Utilizing hypercubic symmetry of the lattice $G(n)$ can be 
written as a linear combination of the values of the propagator at the
corners of the unit hypercube
\beq \label{inival}
G(0,0,0,0) \ , \ G(1,0,0,0) \ , \ G(1,1,0,0) \ , \ G(1,1,1,0) \ , \ G(1,1,1,1) .
\eeq
The second value $G(1,0,0,0)$ can be eliminated by making use of the identity~(\ref{greendiff}).
In the massless case, two more restrictions on the remaining four values of the
Greens functions can be derived by reducing the relation in Eq.~(\ref{recursion})
to a one-dimensional recursion along the lattice axes and 
constructing two linear combinations of the Greens function
values along the lattice axes which are independent of the lattice site $n$ \cite{LW}.
One can then perform the limit $n \to \infty$ and compare it with the asymptotic form of the 
massless Greens function $G_0$
yielding two more relations for the initial propagator values in (\ref{inival}).
Therefore, out of the five initial values three can be eliminated and the entire massless 
Greens function can be expressed in terms of just two values, say $G_0(0,0,0,0)$ and $G_0(1,1,0,0)$, which
have been calculated by L\"uscher and Weisz to very high precision.

For general non-vanishing mass, on the other hand, linear combinations of the propagator values
along the axes which are independent of the lattice site cannot be constructed.
(An exception to this are  the imaginary masses $m^2 = -4 , -8, -12$.
For a similar observation in two dimensions cf. \cite{M}.)
Therefore, with the method described above the initial values of the Greens function 
close to the origin cannot be reduced further by employing its asymptotic form.

Nonetheless, there exists an additional constraint for the massive Greens function which 
allows us to express its values at any lattice site in terms of the massless
Greens function $G_0$ and just two additional constants which are independent of the mass.
As we will see, these additional two coefficients can be calculated to very high precision 
yielding a precise determination of the Greens function 
for $m < 2$.
This constraint is based on the identity for the modified Bessel functions $I_n$ \cite{G}
\beq \label{bessel}
\lambda \ [ I_{n-1} (\lambda) - I_{n+1} (\lambda) ] = 2 n I_{n} (\lambda) \ .
\eeq
The Greens function can be written as an integral over the Bessel functions $I_n$
\beq  \label{intrepr}
G(n_1, n_2, n_3, n_4) = \frac{1}{2} \int_0^\infty d \lambda \ e^{- m^2 \lambda /2 - 4 \lambda}
        I_{n_1} (\lambda) I_{n_2} (\lambda) I_{n_3} (\lambda) I_{n_4} (\lambda) \ ,
\eeq
and utilizing the identity (\ref{bessel}) we obtain the relation
\beqa  \label{greenint}
G(n_1+1, n_2, n_3, n_4) &=& G(n_1-1, n_2, n_3, n_4) + G_0(n_1+1, n_2, n_3, n_4) \no \\
&-&  G_0(n_1-1, n_2, n_3, n_4)  + n_1 \int_0^{m^2} \! \! \! d \mu^2 \ G(n_1, n_2, n_3, n_4; \mu^2) \ ,
\eeqa
where $\mu$ is the mass of the Greens function in the integral and $G_0$
  is the massless Greens function.
Therefore, knowledge of the functional dependence of $G$ on the mass at lower lattice 
sites yields the mass-dependence at higher lattice sites.
This constraint in combination with the recursion relation (\ref{recursion}) 
allows us to calculate $G(1,1,0,0),  G(1,1,1,0),  G(1,1,1,1)$
starting from the Greens function at the origin, the tadpole $G(0,0,0,0)$, if we make use of its functional
dependence on $m$. As a matter of fact, it is well known that the tadpole can be expressed
by the following small mass expansion~\cite{Luscher,Symanzik}
\beq \label{tadpole}
G(0,0,0,0) = \sum_{i=0}^\infty a_i \ m^{2i} + m^2  \ln  m^2 \ \sum_{i=0}^\infty b_i \ m^{2i}
\eeq
with expansion coefficients $a_i$ and $b_i$.
We will show below that this expansion converges absolutely for masses $m^2 <4$.
Employing this expression, one uses then the relation
\beq
G(2, 0, 0, 0) = G(0, 0, 0, 0)  + G_0(2, 0, 0, 0) 
-  G_0(0, 0, 0, 0)  + \int_0^{m^2} \! \! \! d \mu^2 \ G(1, 0, 0, 0; \mu^2) \ ,
\eeq
where $G(1, 0, 0, 0)$ can be directly derived from $G(0, 0, 0, 0)$ via Eq.~(\ref{greendiff})
and is easily integrated utilizing
\beq
\int_0^{m^2}\mu^{i}d\mu=\frac{m^{2(i+1)}}{i+1}  \qquad  \mbox{and} \qquad 
\int_0^{m^2}\mu^{i}\ln \mu \ d\mu=\frac{m^{2(i+1)}}{(i+1)^2}  \Big[(1+i)\ln m^2 -1\Big] \ .
\eeq
Next, the recursion relation (\ref{recursion}) can be employed to obtain  $G(1, 1, 0, 0)$ 
\beq
G(1, 1, 0, 0) =  \frac{1}{6} \Big( [8 + m^2] \ G(1, 0, 0, 0) - G(2, 0, 0, 0) - G(0, 0, 0, 0)  \Big) \ .
\eeq
Similar steps can be taken to obtain $G(2, 1, 0, 0)$
\beq
G(2, 1, 0, 0) = G(1, 0, 0, 0)  + G_0(2, 1, 0, 0) 
-  G_0(1, 0, 0, 0)  + \int_0^{m^2} \! \! \! d \mu^2 \ G(1, 1, 0, 0; \mu^2) 
\eeq
which in turn leads to
\beq
G(1, 1, 1, 0) =  \frac{1}{2} \Big( \Big[4 + \frac{m^2}{2} \Big] \ G(1, 1, 0, 0) 
                   - G(2, 1, 0, 0) - G(1, 0, 0, 0)  \Big) \ .
\eeq
Finally, one has
\beq
G(2, 1, 1, 0) = G(1, 1, 0, 0)  + G_0(2, 1, 1, 0) 
-  G_0(1, 1, 0, 0)  + \int_0^{m^2} \! \! \! d \mu^2 \ G(1, 1, 1, 0; \mu^2) 
\eeq
and
\beq
G(1, 1, 1, 1) =  \frac{1}{2} \Big( [8 + m^2] \ G(1, 1, 1, 0) 
                   - 3 \ G(2, 1, 1, 0) - 3 \ G(1, 1, 0, 0)  \Big) \ .
\eeq
We have thus calculated the Greens function at four lattice sites of the
unit hypercube $G(1,0,0,0)$,  $G(1,1,0,0)$,  $G(1,1,1,0)$,  $G(1,1,1,1)$
recursively by making use of Eqs.~(\ref{greendiff}), (\ref{recursion}) and (\ref{greenint}) which is
sufficient to calculate $G(n_1,n_2,n_3,n_4)$ at all other lattice sites. 
As input we merely need the massless Greens function which is already known to high precision
and the Greens function at the origin, i.e. the tadpole $G(0,0,0,0)$.

For the calculation of the massive Greens function
it is therefore sufficient to determine the coefficients $a_i$ and $b_i$ 
in Eq.~(\ref{tadpole}) to very high precision.
Moreover, one can express all expansion coefficients $a_i$ and $b_i$ just in terms of the leading
coefficients $a_0, a_1, a_2$ and $b_0$. Since $a_0$ is the massless tadpole
$G_0(0,0,0,0)$ and $b_0$ can be calculated exactly, one is then left with the task
to determine the constants $a_1, a_2$ very precisely.
To this end, consider the differential equation for the modified Bessel function $I_n(\lambda)$
\beq
\lambda^2 \frac{d^2}{d \lambda^2} I_n(\lambda) + \lambda \frac{d}{d \lambda} I_n(\lambda) 
 - (\lambda^2 + n^2) \, I_n(\lambda) = 0 \ .
\eeq
It follows that the product $I_0(\lambda)^4$ of Bessel functions satisfies the
differential equation
\beqa
&& \bigg( \lambda^4 \frac{d^5}{d \lambda^5} + 10 \lambda^3 \frac{d^4}{d \lambda^4} 
   - 5 \lambda^2 (4 \lambda^2 -5  ) \frac{d^3}{d \lambda^3}     
   - 15 \lambda (8 \lambda^2 -1  ) \frac{d^2}{d \lambda^2}  \no \\ 
 && \ \  +  (64 \lambda^4 - 152 \lambda^2  +1  ) \frac{d}{d \lambda}   
   + 32 \lambda (4 \lambda^2 -1  )   \bigg) I_0(\lambda)^4 = 0 \ .
\eeqa
With the representation
\beq
G(0,0,0,0) = \frac{1}{2} \int_0^\infty d \lambda \ e^{- m^2 \lambda /2 - 4 \lambda}
        I_{0} (\lambda)^4 \ ,
\eeq
this translates into a differential equation for the tadpole~\cite{Joyce} 
\beqa  \label{diffeqtad}
 && \bigg( ( 8 t^5 - 160 t^3 + 512 t) \frac{d^4}{d(m^2)^4} 
       + ( 40 t^4 - 480 t^2 + 512) \frac{d^3}{d(m^2)^3} \no \\
  &&   \ \      +  ( 50 t^3 - 304 t) \frac{d^2}{d(m^2)^2} 
       +   ( 15 t^2 - 32 ) \frac{d}{d(m^2)} + \frac{t}{2} \bigg) G(0,0,0,0) = 0 \ ,
\eeqa
where $t = 4 + m^2/2$.
Inserting the ansatz (\ref{tadpole}) into Eq.~(\ref{diffeqtad}) each
power $m^{2i}$ and  $m^{2i}\ln m^2$ must vanish separately. This leads to recursive relations
between the expansion coefficients $a_i, b_i$ which are then expressible 
as linear functions of the leading coefficients $a_0, a_1, a_2$ and
$b_0$. The relations between the coefficients $b_{0,1,2,3}$  
are given by
\beqa \label{b0b2}
&& 384 b_0 + 3072 b_1 = 0 \ , \no \\
&& 208 b_0+ 3968 b_1 + 18432 b_2 = 0 \ , \no \\
&& 62 b_0 + 2512 b_1 + 31104 b_2 + 110592 b_3 = 0 \ ,
\eeqa
while for the higher coefficients they read
\beqa  \label{recursionb}
&&i^4 b_{i-2}+(8 + 40 i + 80 i^2 + 80 i^3 + 40 i^4)b_{i-1} +
(832 + 2784 i + 3632 i^2 + 2240 i^3 + 560 i^4) b_i\no \\
&+& (15872 + 43008 i + 43136 i^2 + 19200 i^3 + 3200 i^4) b_{i+1} + (73728 +
172032 i \no\\
&+& 141312 i^2 + 49152 i^3 + 6144 i^4) b_{i+2}=0 \label{bi}
\eeqa
with $i\ge 2$. 
The relations between the coefficients $a_i$ and $b_i$ are 
\beq
2 a_{0} + 208 a_{1} + 3968 a_{2} + 18432 a_{3} + 696 b_{0} + 10752 b_{1} + 
  43008 b_{2} = 0,
\eeq
\beq
a_{0} + 248 a_{1} + 10048 a_{2} + 124416 a_{3} + 442368 a_{4} + 600 b_{0} + 
  19008 b_{1} + 199680 b_{2} + 626688 b_{3} = 0,
\eeq
and for $i\ge 2$ 
\beqa \label{recursiona}
&&i^4 a_{i-1} + (8 + 40 i + 80 i^2 + 80 i^3 + 40 i^4) a_{i} + 
(832 + 2784 i + 3632 i^2 + 2240 i^3 + 560 i^4) a_{i+1} \no\\
&+&(15872 + 43008 i + 43136 i^2 + 19200 i^3 + 3200 i^4) a_{i+2} + 
(73728 + 172032 i + 141312 i^2 \no\\
&+& 49152 i^3 + 6144 i^4) a_{i+3} + 
4 i^3 b_{i-2} + (40 + 160 i + 240 i^2 + 160 i^3) b_{i-1} + 
(2784 + 7264 i \no\\
&+& 6720 i^2 + 2240 i^3) b_{i} + 
(43008 + 86272 i + 57600 i^2 + 12800 i^3) b_{i+1} \no\\
&+&(172032 + 282624 i + 147456 i^2 + 24576 i^3) b_{i+2} = 0.
\eeqa
The constant $a_0$ has already been given  in \cite{CAP} with a
 precision of $10^{-396}$
\beq  \label{r-0}
a_0 = 0.154933390231060214084837208107375088769161133645219\dots ,
\eeq
and we are left with the precise determination of $a_1$ and $a_2$.
This can be accomplished by splitting the tadpole integral $G(0,0,0,0)$ into two parts~\cite{Joyce}
\beq  \label{splitint}
G(0,0,0,0) = \frac{1}{2} \int_0^\Lambda d \lambda \ e^{- m^2 \lambda /2 - 4 \lambda}
        I_{0} (\lambda)^4 + \frac{1}{2} \int_\Lambda^\infty d \lambda \ e^{- m^2 \lambda /2 - 4 \lambda}
        I_{0} (\lambda)^4  \ ,
\eeq
where we leave the parameter $\Lambda$ unspecified for the time being.
The first integral is analytic in the mass and can be expanded in $m^2$, 
while the logarithmic pieces proportional to $\ln m^2$
are contained in the second integral. The integration limit $\Lambda$ 
is now chosen large enough, say $\Lambda =200 \sim 300$,
such that the integrand in the second integral can be replaced by the asymptotic form
for the Bessel functions at large arguments. Hence we obtain 
\beq  \label{numcoeff}
G(0,0,0,0) = \frac{1}{2} \int_0^\Lambda d \lambda \ e^{- m^2 \lambda /2 - 4 \lambda}
        I_{0} (\lambda)^4 + \frac{1}{2} \sum_{n=2}^\infty c_n
        \int_\Lambda^\infty d \lambda \ e^{- m^2 \lambda /2 }  \lambda^{-n}  
\eeq
with well-known expansion coefficients $c_n$. The second integration can be performed
analytically, while the first integral can be evaluated numerically to very high precision,
e.g., by employing a Gau{\ss}-Legendre algorithm~\footnote{Since the
  integrand $\ e^{- m^2 \lambda /2 - 4 \lambda} I_{0} (\lambda)^4 $ is an
  analytic function in $\lambda$, the interpolation procedure converges.}. 
  Although the two single integrals
depend on $\Lambda$, the dependence cancels out in the sum of both terms.
Equation~(\ref{numcoeff}) is therefore suited to determine the constants $a_0, a_1, a_2$ and $b_0$.
As a check we first calculate $a_0$
\beq
a_0 = G_0(0,0,0,0) =  \frac{1}{2} \int_0^\Lambda d \lambda \ e^{- 4 \lambda}
        I_{0} (\lambda)^4 + \frac{1}{2} \sum_{n=2}^\infty \frac{c_n}{(n-1) \Lambda^{n-1}}
\eeq
and obtain agreement with the value given in Eq.~(\ref{r-0}) up to a precision of $10^{-150}$.
This precision is sufficient for our purposes, although it could be easily improved further
by using slightly more computer time. The direct integration proposed here
is thus an alternative method to the one advocated in \cite{LW} 
for obtaining the massless Greens function at the origin to very high precision.
For the coefficients $a_1$ and $a_2$ this procedure yields~\footnote{The first nine digits of $a_1$ 
can be found, e.g., in~\cite{MontMu}}
\beqa
a_1 &=& - \frac{1}{4} \int_0^\Lambda d \lambda \ \lambda \ e^{- 4 \lambda}
        I_{0} (\lambda)^4 + \frac{1}{2}\sum_{n=2}^\infty 
	\frac{c_n}{\Lambda^{n-1}} \ T_1 \left\{ E_n\left(\frac{\Lambda m^2}{2}\right) \right\} \no \\[0.2cm]
    &=&-0.030345755097111005404345333329331191871789937096381870985508127022
\no\\
&&928862366546240927128051477698201507815845611452036369186167716864948\no\\
&&113458657909428\dots, \\[0.3cm]
a_2 &=&  \frac{1}{16} \int_0^\Lambda d \lambda \ \lambda^2 \ e^{- 4 \lambda}
        I_{0} (\lambda)^4 + \frac{1}{2}\sum_{n=2}^\infty 
	\frac{c_n}{\Lambda^{n-1}} \ T_2\left\{ E_n\left(\frac{\Lambda m^2}{2}\right)\right\} \no \\[0.2cm]
    &=&0.0027759274572839795869710267307723922239912302615036308120948317995
\no\\
&&820137296629367446462391602115470725514438235848939713163742942230032\no\\
&&62761367623512\dots,
\eeqa
where $E_n$ is the exponential integral function defined by
\beq
E_n(z)=\int_1^\infty \frac{\exp(-z t)}{t^n}dt,
\eeq
and $T_i$ denotes the $i$-th coefficient in the asymptotic expansion of $E_n$  in 
$m^2$.
Utilizing the series representation of the exponential integral function
\beq
E_n(z)=\frac{(-z)^{n-1}}{(n-1)!}\left[- \ln z+\psi(n)\right]
-\sum_{m=0,m\neq n-1}^\infty\frac{(-z)^m}{(m-n+1)m!} \ ,
\eeq
where $\psi(n)$ is digamma function defined by
\beq
\psi(1)=-\gamma,\quad \psi(n)=-\gamma + \sum_{m=1}^{n-1}\frac{1}{m},
\eeq
one can immediately read off the coefficients $T_i$ 
\beqa
\frac{1}{2}\sum_{n=2}^\infty 
	\frac{c_n}{\Lambda^{n-1}} \ T_1 \left\{ E_n\left(\frac{\Lambda m^2}{2}\right)\right\}&=&
-\frac{1}{4}\left[c_2\left\{\psi(2)-\ln\left(\frac{\Lambda}{2}\right)\right\} -\sum_{n=3}^\infty c_n\frac{\Lambda^{2-n}}{2-n}\right],\\
\frac{1}{2}\sum_{n=2}^\infty 
	\frac{c_n}{\Lambda^{n-1}} \ T_2 \left\{ E_n\left(\frac{\Lambda m^2}{2}\right)\right\}&=&
\frac{1}{16}\left[c_3\left\{\psi(3)-\ln\left(\frac{\Lambda}{2}\right)\right\}
- c_2 \Lambda - \sum_{n=4}^\infty
c_n\frac{\Lambda^{3-n}}{3-n}\right].
\eeqa
We have varied $\Lambda$ within the range $\Lambda = 200 \ldots 400$ and confirmed
that the results are indeed independent of $\Lambda$ up to a precision of $10^{-150}$.
The parameter $b_0$, on the other hand, originates solely from the second integral
and can be calculated exactly: $b_0 = 1/16\pi^2$. From Eqs.~(\ref{b0b2}-\ref{bi}) 
it follows then that all coefficients $b_i$ can be calculated analytically.
Having determined the constants $a_0,a_1$ and $a_2$ to very high precision,
we use the recursion relations for the coefficients $a_i$ and $b_i$ which
follow from Eq.~(\ref{diffeqtad}) to calculate the last remaining value of the
Greens function on the unit hypercube, $G(0,0,0,0)$.
As an additional check, we have calculated the Greens function values 
$G(0,0,0,0), G(1,1,0,0), G(1,1,1,0), G(1,1,1,1)$
by splitting the integral representation of Eq.~(\ref{intrepr})
into two ranges $[0,\Lambda]$ and $[\Lambda,\infty)$  analogous to Eq.~(\ref{splitint})
and evaluating the integrals directly. The results obtained in this way agree
with the ones achieved by applying the recursion relation up to $10^{-135}$.
With the method outlined above we thus confirm the form of the tadpole in Eq.~(\ref{tadpole}).

Employing the recursion relations Eq.~(\ref{recursionb}) and Eq.~(\ref{recursiona}) 
we now illustrate that the small mass expansion of the tadpole (\ref{tadpole})
converges absolutely for $m^2 < 4$. For large enough $i$ the recursion relation for the $b_i$
simplifies to
\beq   \label{asymrecb}
\frac{b_{i+1}}{b_i} = - \frac{25}{48} - \frac{b_{i-3}}{6144 \, b_i}- \frac{5\, b_{i-2}}{768 \, b_i}
             - \frac{35\, b_{i-1}}{384 \, b_i} \ .
\eeq
If we assume that for large $i$ the ratio $b_{i+1}/b_i$ converges asymptotically towards
a constant value, then Eq.~(\ref{asymrecb}) has the solutions $b_{i+1}/b_i = - \frac{1}{4},
- \frac{1}{8}, - \frac{1}{12}, - \frac{1}{16}$. We have confirmed numerically that
the ratio tends indeed towards $-\frac{1}{4}$. This implies that for large $i$ the parameters $b_i$
scale as $|b_i| \propto 4^{-i}$. The logarithmic part in Eq.~(\ref{tadpole}) converges
thus absolutely for $m^2<4$.

Assuming also a constant ratio $a_{i+1}/a_i$ for large enough $i$ and
the boundedness of $b_{i}/a_i$ we arrive at the same simplified recursion relation
for the $a_i$
\beq   \label{asymreca}
\frac{a_{i+1}}{a_i} = - \frac{25}{48} - \frac{a_{i-3}}{6144 \, a_i}- \frac{5\, a_{i-2}}{768 \, a_i}
             - \frac{35\, a_{i-1}}{384 \, a_i} \ .
\eeq
Again, we have checked numerically that both the ratio $a_{i+1}/a_i$ tends towards $-\frac{1}{4}$
and $b_{i}/a_i$ is bounded. This implies that the polynomial part of the tadpole
in (\ref{tadpole}) and hence the entire small mass expansion of the tadpole converges
absolutely for $m^2<4$.

\begin{figure}
\begin{center}
\includegraphics[height=8cm,angle=-90]{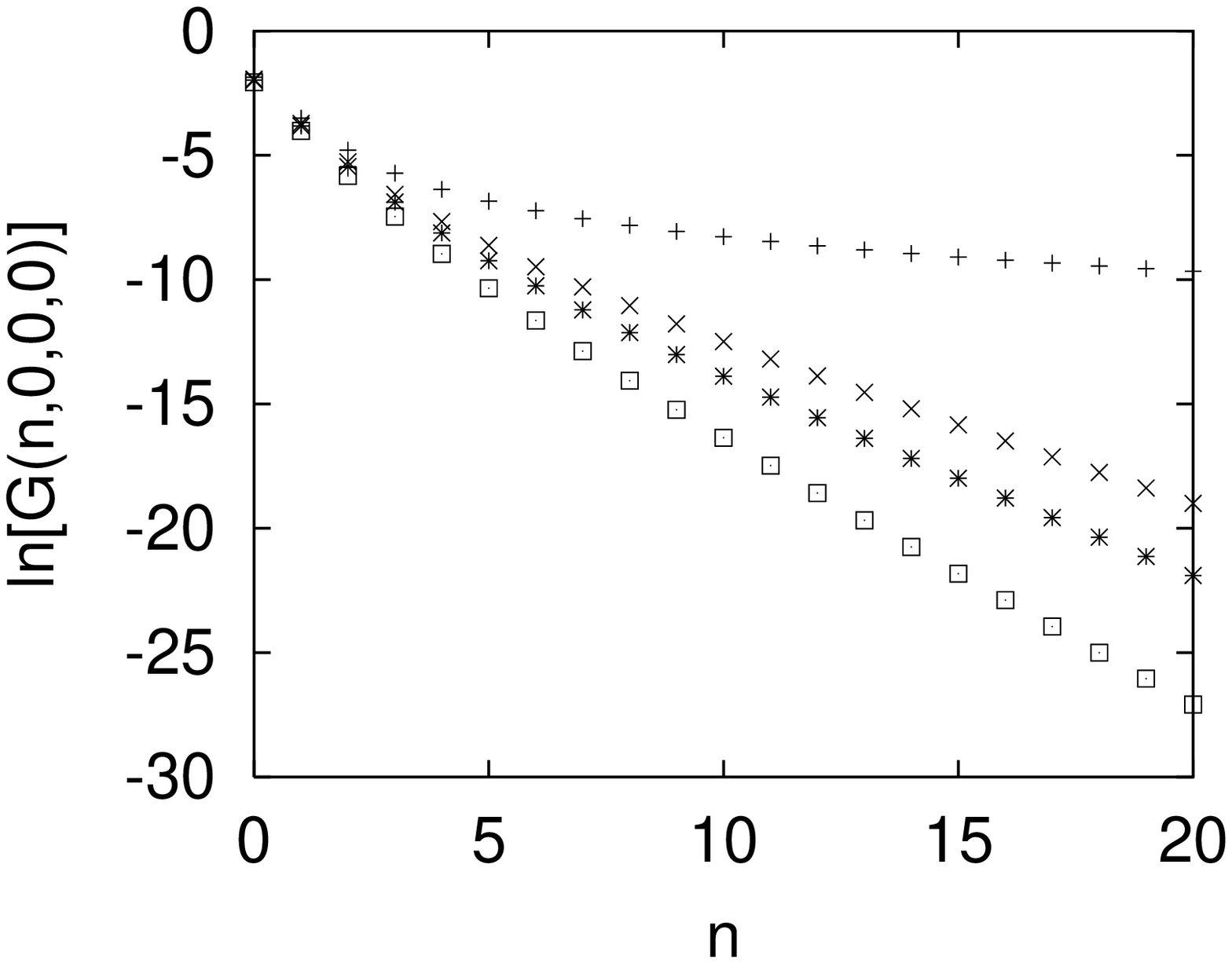}
\includegraphics[height=8cm,angle=-90]{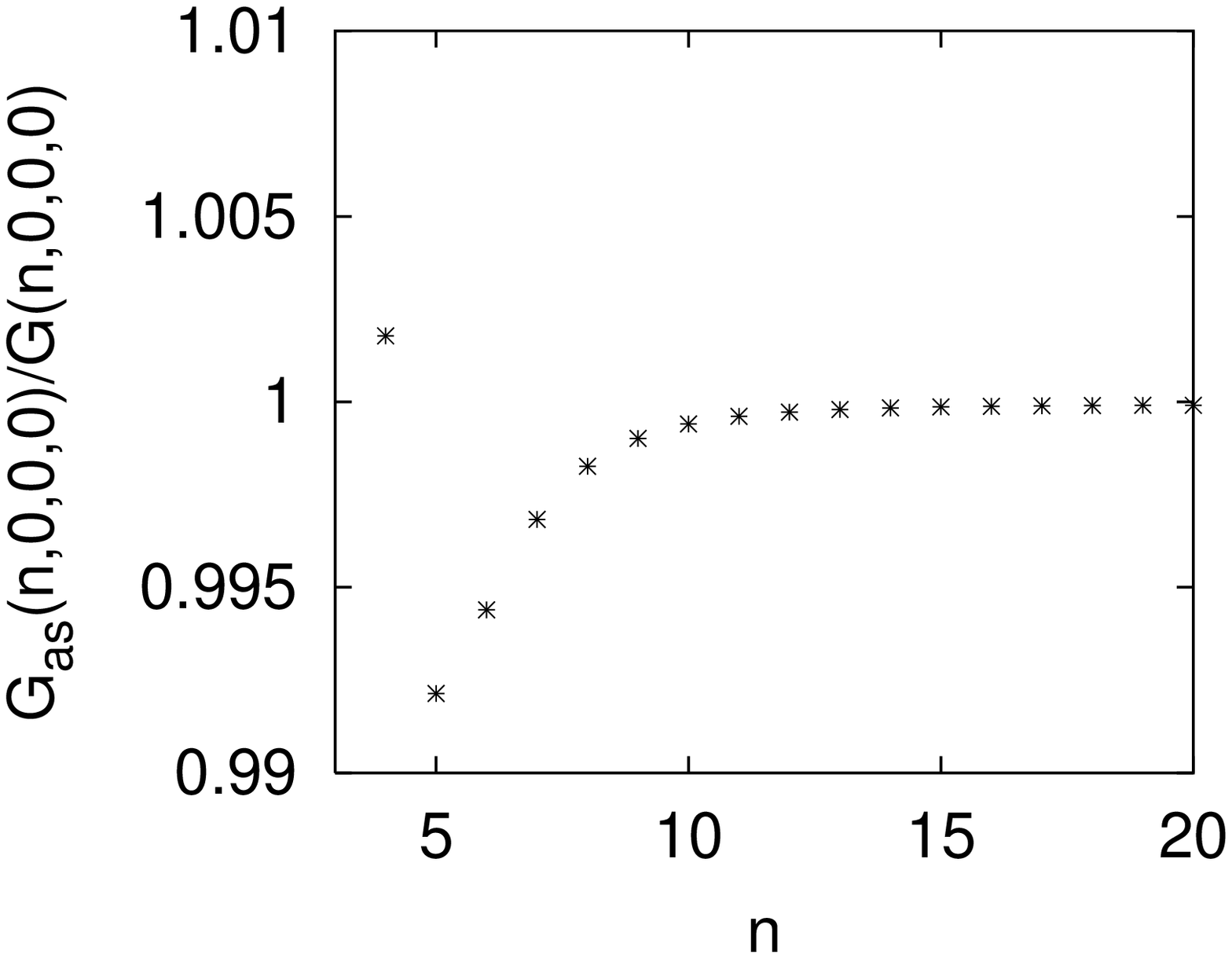}
\end{center}
\caption{ Left panel: Shown are the logarithms of the Greens functions $G(n,0,0,0)$
for masses $m^2 = 0,\frac{3}{10},\frac{1}{2},1$ (from top to bottom). Right panel: Ratio of 
$G_{as}(n,0,0,0)/G(n,0,0,0)$ for $m^2=\frac{1}{2}$. \label{GreenfM}}
\end{figure}

Next, we would like to confirm the results for the asymptotic behavior of the lattice Greens function 
for large $|n|=\sqrt{n_1^2+n_2^2+n_3^2+n_4^2}$. In~\cite{Paladini} Paladini and Sexton 
have derived an asymptotic form of $G(n)$ for large $|n|$, $G_{as}(n)$, in terms of modified Bessel 
functions of the second kind, $K_l(m |n|)$. 
In the left panel of Fig.~\ref{GreenfM} we have plotted our results  $\ln[G(n,0,0,0)]$
for four different masses. We observe that with increasing $n$ \
$\ln[G(n,0,0,0)]$ becomes an asymptotically linear function in~$n$.
As for large $n$ the Bessel 
functions $K_l(m n)$ decrease exponentially as $K_l(m n)\sim \sqrt{\frac{\pi}{2 m\,n}}\exp(-m\,n)$, 
the different slopes in the plot reflect the different masses.
In the right panel, the ratio $G_{as}(n,0,0,0)/G(n,0,0,0)$ is plotted for $m^2=1/2$,
demonstrating clearly the fast convergence of the asymptotic expansion towards
the exact result.

So far, we have discussed the case $m < 2$. The case $m \gtrsim 1$, on the other
hand, can be calculated in a straightforward manner. Starting from Eq.~(\ref{intrepr}), we replace 
the Bessel functions $I_\nu$ by their Taylor series to obtain
\beq  \label{intbigmass}
G(n_1, n_2, n_3, n_4) = \frac{1}{2} \int_0^\infty d \lambda \ e^{- m^2 \lambda /2 - 4 \lambda}
        \sum_{i=0}^{\infty} d_i (n_1, n_2, n_3, n_4) \, \lambda^i ,
\eeq
where the expansion coefficients $d_i$ are derived from the product of the Taylor series of the $I_\nu$.
Since the integrand in Eq.~(\ref{intbigmass}) is absolutely convergent,
summation and integration can be interchanged
leading to
\beq  \label{massum}
G(n_1, n_2, n_3, n_4) = \frac{1}{2} \sum_{i=0}^{\infty} d_i (n_1, n_2, n_3, n_4) \,   \frac{i!}{4 ^{(i+1)}} 
                       \left( 1 + \frac{m^2}{8} \right)^{-(i+1)} .
\eeq
We have checked numerically that this series converges sufficiently fast for masses $m \gtrsim 1$.
E.g., for $m=1$ we obtain a precision of $10^{-160}$ by summing up the first $n=7000$ 
terms in (\ref{massum}),
while for $m^2=8$ the precision is $10^{-191}$ when taking the first 700 terms into account.
Note, that the convergence behavior of the sum in Eq.~(\ref{massum}) worsens with decreasing
$m$ and is extremely inefficient for $m \ll 1$. It is therefore not suited to perform the
limit $m \to 0$.

To conclude, we have presented the precise determination
of the Greens function of a massive scalar field on the lattice which is based on
a recursion relation for the values of the Greens function values at different lattice sites
similar to the one obtained in the massless case \cite{LW}. 
We have shown that the Greens function for masses $m < 2$ is fixed by the knowledge of the 
massless Greens function and only two additional mass-independent
constants $a_1$ and $a_2$, which can be determined to very high 
precision. We have also confirmed the usefulness of the large $|n|$ asymptotic expansion of $G(n)$ 
given by Paladini and Sexton~\cite{Paladini} and demonstrated its fast convergence towards
the exact result. 
For $m \gtrsim 1$ the Greens function can be calculated directly by writing the
Greens function as an integral over the Bessel functions and replacing these by the corresponding
Taylor series.

With minimal amount
of computer time we achieve precisions of $10^{-150}$ for the Greens function
values at the origin.  Such precise knowledge of the Greens function is expected to be 
sufficient for carrying out two- and higher-loop calculations. 
The applicability to higher-loop integrals is currently under investigation and could prove useful, e.g.,
in the recently proposed framework of lattice regularized chiral perturbation theory \cite{latchpt}.\\

This work has been partially supported by the Deutsche Forschungsgemeinschaft.


\end{document}